\documentclass[runningheads]{llncs}

\usepackage{cite}
\usepackage[T1]{fontenc}
\usepackage{algorithmic}
\usepackage{graphicx}
\usepackage{subcaption}
\usepackage{textcomp}
\usepackage{xcolor}
\usepackage[inline,shortlabels]{enumitem}

\usepackage{todonotes}

\newenvironment{inlinelist}[1][\roman]{\begin{enumerate*}[label=\emph{(#1*)}]}{\end{enumerate*}}

\usepackage{subfig}
\usepackage{comment}
\usepackage{booktabs}
    
\begin{document}

\newcommand{\sched}{\textsc{DIWS}}
\newcommand{\easybf}{\textsc{EasyBF}}

\title{Duration-Informed Workload Scheduler}

\author{Daniela Loreti\inst{1}\orcidID{0000-0002-6507-7565} \and
Davide Leone\inst{1}\orcidID{0009-0000-7697-8863} \and
Andrea Borghesi\inst{1}\orcidID{0000-0002-2298-2944}}

\authorrunning{Daniela Loreti et al.}

\institute{DISI, University of Bologna, Bologna, Viale Risorgimento 2, Italy\\
\email{\{daniela.loreti,andrea.borghesi3\}@unibo.it}\\
\email{\{davide.leone\}@studio.unibo.it}}

\maketitle

\begin{abstract}
High-performance computing systems are complex machines whose behaviour is governed by the correct functioning of its many subsystems. Among these, the workload scheduler has a crucial impact on the timely execution of the jobs continuously submitted to the computing resources. Making high-quality scheduling decisions is contingent on knowing the duration of submitted jobs before their execution--a non-trivial task for users that can be tackled with Machine Learning. 

In this work, we devise a workload scheduler enhanced with a duration prediction module built via Machine Learning. We evaluate its effectiveness and show its performance using workload traces from a Tier-0 supercomputer, demonstrating a decrease in mean waiting time across all jobs of around 11\%. Lower waiting times are directly connected to better quality of service from the users' point of view and higher turnaround from the system's perspective. 

\keywords{High-Performance Computing, Duration Prediction, Machine Learning}
\end{abstract}

\section{Introduction}\label{sec:intro}
The ever-increasing capabilities of modern High-Performance Computing (HPC) facilities are already going beyond the exascale. Despite this impressive growth, the increase in service demand due to emerging fields of science 
makes the computing power offered by HPC infrastructures a coveted resource. By now, techniques to improve the efficient usage of HPC resources are a pressing need.

Workload schedulers are among the mechanisms that have the greatest impact on HPC facilities' efficiency (as they have on computing systems of any scale). 
Besides, the quality of the scheduling choices has a direct effect on HPC users' experience because it influences the turnaround time of the launched jobs. 
%
The ability of these systems to make high-quality scheduling decisions depends on a variety of factors, not least the availability of reliable estimations of the execution time for each job before it is submitted.
Unfortunately, the ever-growing complexity of parallel HPC software makes performance prediction harder and harder with traditional methods (e.g., with techniques based on code analysis). 

Over the last decades, Machine Learning models have been applied to almost any scientific field in order to improve the quality of predictions and help manage the complexity of domains characterized by many input features.
In particular, although ML techniques have proven successful in the runtime estimation of specific tasks \cite{hutter2014algorithm}, the efficacy of integrating their predictive power in HPC scheduling systems still needs to be explored.

In this work, we focus on ML techniques as a means to manage the complexity of the runtime prediction task for HPC jobs, and we explore the advantages and drawbacks of integrating ML-based runtime estimations into an HPC scheduling policy. 

In detail, the contribution of this paper is two-fold:



\begin{itemize}
    \item Revealing how ML models can be used to predict the duration of the workload in modern, production supercomputers; a thorough analysis of the quality of the estimate is provided using a real supercomputer, Marconi100, as a case study.
    \item Developing an HPC workload scheduler that is informed by the predictions made by the ML models. The scheduler has been validated using an off-the-shelf HPC simulator, demonstrating significant improvements in terms of waiting time (a decrease of 11.21\% with respect to using the standard duration estimate provided by users), mean turnaround time (decreased by 4.35\%) and average job slowdown (decreased by 94.96\%).   
\end{itemize}

Therefore, the rest of the paper is structured as follows. Section \ref{sec:rel} presents an overview of the state of the arts of runtime estimation as well as workload scheduling with a particular focus on HPC environments. Section \ref{sec:time_prediction} evaluates the performance of four different ML methods in predicting the duration of HPC jobs given only features that can be available at submission time as input. The logic of the proposed Duration-Informed Workload Scheduler (\sched{}) is then presented in section \ref{sec:sched} along with a comparison of its performance with a widely employed scheduling policy. Conclusion and future work follow.

\section{Related Works}\label{sec:rel}
\subsection{Runtime Prediction}
\label{sec:rel_runtime}

In recent literature, various machine learning techniques have been used to estimate application workflows, such as decision trees \cite{miu2012predicting} and neural networks \cite{nadeem2017modeling}. A comprehensive survey on resource provisioning prediction models is provided in \cite{amiri2017survey}. It must be noted that the majority of existing works do not explicitly focus on the duration of HPC application as a prediction target. While most studies focus on lightweight algorithms on standard hardware, runtime prediction for HPC workloads remains under-explored; see \cite{pittino2019prediction,de2024machine} for some preliminary results.

\subsection{HPC Workload Scheduling}
\label{sec:rel_sched}

First-Come First-Served (FCFS) and Earliest Available Start Time Yielding (EASY) backfilling are still the most widely used scheduling policies in production HPC systems as they are easy to implement and known to produce good results.
EASY backfilling allows smaller jobs to skip ahead of larger jobs, as long as this does not delay the job at the head of the queue. As such, this policy strongly relies on user-provided runtime estimates, which are known to be significantly inaccurate on average \cite{990753}.
Therefore, some recent works focused on improving EASY backfilling with ML-enhanced runtime predictions \cite{DBLP:conf/xsede/TanashDAHYO19,DBLP:journals/tjs/LiZHJDH21}. These works are limited as they do not consider full Tier-0 systems (but rather focus on smaller-scale datacenters), with a more diverse distribution of job durations.

\section{Runtime Prediction}\label{sec:time_prediction}


In this section, we start by introducing the ML models used to estimate the workload duration on the target supercomputer. Then, we report their performance in terms of prediction accuracy.

\subsection{Prediction Models}\label{sec:time_prediction_result_models}

The ML models selected are the following: 
\begin{inlinelist}
\item Decision Tree Regressor (DT)\cite{breiman1984classification} -- a supervised learning algorithm that constructs a tree-structured model to represent decisions and their consequences. Regression trees partition the input feature space recursively into smaller regions, assigning a numerical value as the output for each region.

\item Random Forest (RF)\cite{breiman2001random} -- an extension of the DT model designed for regression tasks, where the target is a continuous numeric value. It combines predictions from an ensemble of decision trees to improve robustness and accuracy.

\item Gradient Boosting (GB)\cite{friedman2001greedy} -- an ensemble learning technique that incrementally builds a strong predictive model by combining multiple weak learners, often decision or regression trees. Each new model corrects the errors of the previous ones, enhancing predictive performance.

\item Fully Connected Neural Network (FCNN)\cite{bengio2017deep} -- a machine learning model inspired by biological neural networks. It consists of interconnected layers of nodes (neurons), where each layer processes and transforms input data into output predictions. Each neuron's input aggregates outputs from the previous layer, followed by a non-linear transformation to generate the layer's output. In particular, we employed a network with three hidden layers and dropout to prevent overfitting and used the Huber loss, which is less sensitive to outliers. The number of layers and the number of neurons in each layer are the result of a non-exhaustive na\"{i}ve grid-like search, in which we trained a total of 15 networks varying only these two parameters to find the best combination.
\end{inlinelist}

\begin{figure*}
    \centering
    \includegraphics[width=.9\linewidth]{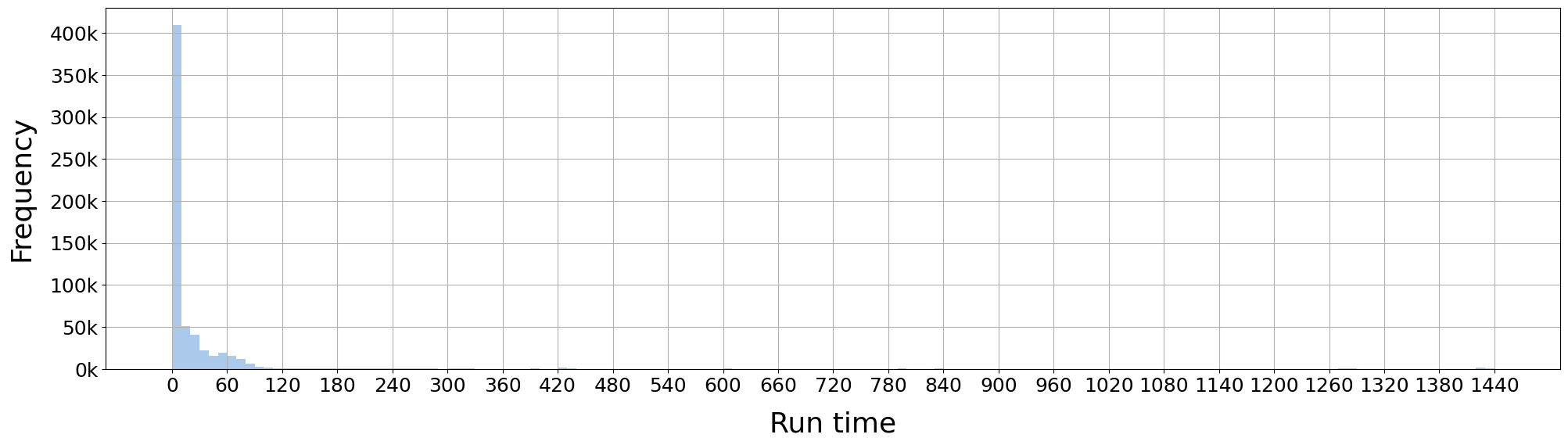}
    \caption{Histogram of the target variable \textit{run\_time}}
    \label{fig:histogram_runtime}
\end{figure*}

\subsection{Empirical Results}\label{sec:time_prediction_result}
We evaluate the performance of the aforementioned ML techniques in predicting the runtimes from PM100 \cite{DBLP:conf/sc/AnticiABK23}, a large dataset of real-life job runs, derived from an accurate elaboration of a two-years-long data collection \cite{borghesi2023m100} from a production supercomputer: MARCONI100 hosted by the HPC center CINECA\footnote{https://www.hpc.cineca.it/systems/hardware/marconi100/}.
The considered dataset consists of 628.977 elements and a set of submission time features for each job, which are described in Table \ref{tab:data_description}. In Table \ref{tab:data_analysis}, the features are shown together with the target variable \textit{run\_time} and a statistical description of each field. We have selected a subset of the whole PM100 dataset (comprising more than one million jobs), as the removed entries contain missing values. The impact of missing data on the accuracy of ML models is an interesting problem by itself, but we leave it for future research, as in this work we want to focus on the base problem of predicting workload duration.

The analysis of the data reveals several significant characteristics. A particularly striking feature is the high variability observed in the \textit{cpu} and \textit{mem(GB)} metrics, as indicated by their large standard deviations and the substantial range between the minimum and maximum values. This variability highlights the heterogeneous nature of the dataset in these dimensions. Additionally, the data exhibits pronounced skewness across most variables. This skewness is primarily driven by the presence of a few extreme outliers, which have a notable impact on the mean values. Such outliers inflate the averages, creating a substantial gap between the mean and the more representative median values. The consistently lower median values across most variables suggest that the majority of the dataset is concentrated around lower values, while a small number of high-value entries significantly raise the mean. These trends are further corroborated by the visual analysis of histograms (as the one showed in Fig. \ref{fig:histogram_runtime}), where the skewness and the influence of outliers are clearly discernible. The histograms provide a compelling illustration of how the distribution of values deviates from symmetry, emphasizing the predominance of lower values juxtaposed with infrequent but extremely high values.
To summarize, we are dealing with a dataset coming from a real Tier-0, production supercomputer; this entails, that the workload we are considering is complex and non-trivial to handle. 

\begin{table*}
    \centering
    \scriptsize
    \begin{tabular}{|l|l|}
        \hline
        \textbf{Feature Name} & \textbf{Description} \\ \hline
        cpu & Number of CPU cores requested by the job. \\ \hline
        mem (GB) & Amount of memory requested by the hob. \\ \hline
        node & Number of nodes requested for the job. \\ \hline
        gres/gpu & GPU resources requested by the job. \\ \hline
        user\_id & Identifier of the user submitting the job. \\ \hline
        qos & Quality of Service level associated with the job. \\ \hline
        time\_limit & Maximum runtime allowed for the job. \\ \hline
    \end{tabular}
    \caption{Brief description of the features.}
    \label{tab:data_description}
\end{table*}

\begin{table*}[]
    \centering
    \scriptsize
    \begin{tabular}{|l|r|r|r|r|r|r|r|r|}
    \hline
         & \textbf{CPU} & \textbf{mem(GB)} & \textbf{nodes} & \textbf{GRES/GPU}& \textbf{user\_id} & \textbf{QoS} & \textbf{time\_limit} & \textbf{run\_time} \\ \hline
         \textbf{mean} & 121.379 & 236.068 & 1.693 & 5.630 & 110.895 & 0.051 & 1038.069 & 43.433\\
         \textbf{std}  & 246.657 & 1008.594 & 6.961 & 27.927 & 118.594 & 0.368 & 506.318 & 168.719\\
         \textbf{min}  & 1.000 & 0.098 & 1.000 & 1.000  & 0.000 & 0.000 & 1.000 & 0.017 \\
         \textbf{25\%} & 4.000 & 7.813 & 1.000 & 1.000 & 2.000 & 0.000 & 720.000 & 0.017 \\
         \textbf{50\%} & 80.000 & 230.000 & 1.000 & 4.000 & 93.000 & 0.000 & 1440.000 & 0.83 \\
         \textbf{75\%} & 128.000 & 237.5000 & 1.000 & 4.000 & 191.000 & 0.000 & 1440.000 & 22.700 \\
         \textbf{max} & 32768.000 & 61500.000 & 256.000 & 1024.000 & 387.000 & 3.000 & 1440.000 & 1439.912 \\ \hline
    \end{tabular}
    \caption{Brief statistical description of the dataset.}
    \label{tab:data_analysis}
\end{table*}

The performance of four ML models—Decision Tree, Random Forest, Gradient Boosting, and Neural Network (see Section~\ref{sec:time_prediction_result_models}) -- was evaluated based on their predictive accuracy and error characteristics. The following metrics were used for comparison: Mean Absolute Error (MAE), Mean Squared Error (MSE), Root Mean Squared Error (RMSE), the coefficient of determination ($R^2$), and the 95\% confidence interval for prediction errors. Furthermore, the analysis included an investigation of error characteristics categorized as overestimations, underestimations, and exact estimations (down to half a second).

Furthermore, we compute the \textit{effectiveness} of the four considered models by measuring the improvements in approximating the actual run time with respect to the \textit{time\_limit} column, which reports the user-provided estimation. As the consequences of an underestimation are--in general--more problematic than overestimation, we also consider \textit{valid predictions} the ones not lower than the actual run time. Table \ref{tab:preproc_results} shows the performance of the four considered models; the data has been normalized using the MinMax scaler algorithm implemented in Scikit-learn for the sake of the neural network (these models are notoriously better performing with normalized data). The dataset was randomly split between training and testing sets, with a split ratio of 70\%/30\%.

\begin{table*}[]
    \centering
    \scriptsize
    \begin{tabular}{|l|c|c|c|c|}
    \hline & \textbf{Decision Tree} & \textbf{Random Forest} & \textbf{Gradient Boosting}  &\textbf{Neural Network} \\ \hline
        \textbf{MAE}& 23.51 & 23.53 & 40.11 & 21.95 \\
        \textbf{MSE}& 8001.99 & 7968.58 &13060.00 &9202.41\\
        \textbf{RMSE}& 89.45 & 89.27 &114.28 &95.93\\
        \textbf{$R^2$}& 0.72 & 0.72 &0.54 &0.68 \\
\textbf{Confidence interval (95\%)} & [0.00, 326.70] &[0.00, 325.60] & [0.00, 227.48] & [0.00, 344.92] \\ \hline
 \multicolumn{5}{|c|}{\textbf{OVERESTIMATION}} \\ \hline
 \textbf{Total cases}  & 79.49\% & 79.59\% & 82.07\% & 80.79\% \\
\textbf{min error }& 0.01 & 0.01 & 0.01 & 0.01\\
\textbf{max error} & 1431.00 & 1303.47 & 806.34 & 1624.70\\
\textbf{avg error} & 14.76 & 14.72 & 24.35 & 12.39 \\
\textbf{error < 60 minutes} & 96.30\% & 96.26\% & 92.38\% & 97.98\% \\ \hline
 \multicolumn{5}{|c|}{\textbf{UNDERESTIMATION}} \\ \hline
 \textbf{Total cases}  & 20.02\% & 19.95\% & 17.93\% & 19.18\% \\
\textbf{min error }& 0.01 & 0.01 & 0.01 & 0.01\\
\textbf{max error} & 1425.53 & 1425.54 & 1418.82 & 1427.30\\
\textbf{avg error} & 58.85 & 59.23 & 112.26 & 62.23 \\
\textbf{error < 60 minutes} & 86.67\% & 86.34\% & 73.95\% & 83.33\% \\ \hline
 \multicolumn{5}{|c|}{\textbf{EXACT ESTIMATION}} \\ \hline
 \textbf{Total cases}  & 0.50\% & 0.47\% & 0.02\% & 0.02\% \\ \hline
 \multicolumn{5}{|c|}{\textbf{EFFECTIVENESS}} \\ \hline
 \textbf{General}  & 78.09\% & 78.23\% & 74.16\% & 78.72\% \\
 \textbf{Valid prediction}  & 97.94\% & 97.96\% & 92.79\% & 97.63\% \\ \hline
    \end{tabular}
    \caption{Performance of the four considered ML models}
    \label{tab:preproc_results}
\end{table*}

From the results RF appears to be the best model, followed at short distances by DT and FCNN, while GB yields the lowest performance. The Neural Network model had the lowest MAE, while RF achieved the lowest MSE,  closely followed by DT. Decision Tree and Random Forest models achieved the highest $R^2$ values, indicating better explanatory power compared to the other models. 
All models predominantly overestimated predictions, with GB showing the highest proportion of overestimations and NN the lowest. Most overestimation errors (above 92\%) were within 60 minutes across models.
GB had the lowest proportion of underestimations -- but the overall performance of GB is lowered by the higher number of overestimates. Exact predictions were rare, with Decision Tree and Random Forest achieving the highest proportion, while GB and FCNN had almost negligible exact predictions (0.02\%).

\subsubsection{Data Augmentation}\label{sec:time_prediction_result_dataAug}

We conducted another experiment to explore the possibility of improving the quality of the predictions. Namely, we performed a data augmentation step before training by adding the average resource requested by each user, i.e., the mean values for the requested number of CPUs, memory, physical nodes, GPUs and time limit. As aspected, the results (shown in Table \ref{tab:dataug_results}) sightly improve with data augmentation.
\begin{table*}[]
    \centering
    \scriptsize
    \begin{tabular}{|l|c|c|c|c|}
    \hline & \textbf{Decision Tree} & \textbf{Random Forest} & \textbf{Gradient Boosting}  &\textbf{Neural Network} \\ \hline
        \textbf{MAE}& 22.24 & 22.26 & 26.01 & 20.53 \\
        \textbf{MSE}& 7312.82 & 7275.61 & 8406.57 & 8623.19\\
        \textbf{RMSE}& 85.52 & 85.30 & 91.69 & 92.86\\
        \textbf{$R^2$}& 0.71 & 0.72 &0.67 &0.66 \\
\textbf{Confidence interval (95\%)} & [0.00, 307.77] &[0.00, 306.92] & [0.00, 291.46] & [0.00, 319.62] \\ \hline
 \multicolumn{5}{|c|}{\textbf{OVERESTIMATION}} \\ \hline
 \textbf{Total cases}  & 79.90\% & 79.98\% & 80.57\% & 79.34\% \\
\textbf{min error }& 0.01 & 0.01 & 0.01 & 0.01\\
\textbf{max error} & 1425.65 & 1425.66 & 1118.25 & 1470.79\\
\textbf{avg error} & 13.97 & 13.97 & 16.20 & 12.26 \\
\textbf{error < 60 minutes} & 96.20\% & 96.15\% & 95.64\% & 97.93\% \\ \hline
 \multicolumn{5}{|c|}{\textbf{UNDERESTIMATION}} \\ \hline
 \textbf{Total cases}  & 19.69\% & 19.63\% & 19.41\% & 20.63\% \\
\textbf{min error }& 0.01 & 0.01 & 0.01 & 0.01\\
\textbf{max error} & 1425.42 & 1425.41 & 1424.76 & 1427.48\\
\textbf{avg error} & 56.25 & 56.48 & 66.72 & 55.36 \\
\textbf{error < 60 minutes} & 87.46\% & 87.29\% & 83.63\% & 86.07\% \\ \hline
 \multicolumn{5}{|c|}{\textbf{EXACT ESTIMATION}} \\ \hline
 \textbf{Total cases}  & 0.41\% & 0.39\% & 0.02\% & 0.03\% \\ \hline
 \multicolumn{5}{|c|}{\textbf{EFFECTIVENESS}} \\ \hline
 \textbf{General}  & 78.81\% & 78.81\% & 78.81\% & 78.81\% \\
 \textbf{Valid prediction}  & 98.51\% & 98.51\% & 98.51\% & 98.51\% \\ \hline
    \end{tabular}
    \caption{Performance of the four considered ML models with data augmentation.}
    \label{tab:dataug_results}
\end{table*}


\subsubsection{Time-consecutive Split Setting}\label{sec:time_prediction_result_online}

Finally, we performed a last experimental evaluation with a different splitting strategy for training and testing sets to simulate a real-life situation better. Since in practice the scheduler typically works on subsequent job submissions, it is requested to estimate the runtime of future jobs given the jobs arrived in the past as training examples. Therefore, randomly splitting the dataset into training and testing sets may not represent a real-life case. In the following, we repeat the evaluations using a consecutive split over time, i.e., all jobs before a certain date are used for training and all those after are used for testing. We chose the date such that the test set contains exactly the same number of jobs as the one with the random split. Table \ref{tab:consec_results} shows that the quality of the results is comparable to those of the previous experiment. In particular, all the error values are better but $R^2$ is worse, indicating that the quality of the models has decreased despite an average better predictive capacity. These results can be explained by the fact that, using the consecutive split, in the test set there are much lower average runtime values than in the training set. Furthermore, the standard deviation of all the test set columns is smaller than that of the training set columns.

\begin{table*}[]
    \centering
    \scriptsize
    \begin{tabular}{|l|c|c|c|c|}
    \hline & \textbf{Decision Tree} & \textbf{Random Forest} & \textbf{Gradient Boosting}  &\textbf{Neural Network} \\ \hline
        \textbf{MAE}& 8.33 & 8.35 & 22.90 & 8.22 \\
        \textbf{MSE}& 3438.32 & 3432.47 & 5086.70 & 3674.63\\
        \textbf{RMSE}& 58.64 & 58.59 & 71.32 & 60.62\\
        \textbf{$R^2$}& 0.62 & 0.62 &0.44 &0.60 \\
\textbf{Confidence interval (95\%)} & [0.00, 155.35] &[152.11] & [0.00, 104.60] & [0.00, 165.33] \\ \hline
 \multicolumn{5}{|c|}{\textbf{OVERESTIMATION}} \\ \hline
 \textbf{Total cases}  & 94.40\% & 94.86\% & 95.99\% & 94.49\% \\
\textbf{min error }& 0.01 & 0.01 & 0.02 & 0.02\\
\textbf{max error} & 1196.12 & 1184.39 & 722.10 & 1311.46\\
\textbf{avg error} & 4.18 & 4.08 & 16.96 & 4.18 \\
\textbf{error < 60 minutes} & 99.44\% & 99.13\% & 98.90\% & 99.36\% \\ \hline
 
 \multicolumn{5}{|c|}{\textbf{UNDERESTIMATION}} \\ \hline
 \textbf{Total cases}  & 5.25\% & 5.13\% & 4.00\% & 5.50\% \\
\textbf{min error }& 0.01 & 0.01 & 0.02 & 0.02\\
\textbf{max error} & 1425.71 & 1425.71 & 1399.33 & 1424.30\\
\textbf{avg error} & 83.43 & 87.44 & 165.44 & 77.58 \\
\textbf{error < 60 minutes} & 81.69\% & 80.87\% & 67.75\% & 82.63\% \\ \hline

 \multicolumn{5}{|c|}{\textbf{EXACT ESTIMATION}} \\ \hline
 \textbf{Total cases}  & 0.35\% & 0.01\% & 0.00\% & 0.01\% \\ \hline
 
 \multicolumn{5}{|c|}{\textbf{EFFECTIVENESS}} \\ \hline
 \textbf{General}  & 94.22\% & 94.27\% & 93.40\% & 93.42\% \\
 \textbf{Valid prediction}  & 99.45\% & 99.37\% & 97.79\% & 98.90\% \\ \hline
    \end{tabular}
    \caption{Performance of the four considered ML models with a consecutive split of train and test sets.}
    \label{tab:consec_results}
\end{table*}

\subsubsection{Discussion}\label{sec:time_prediction_result_discussion}

We notice how in all the tested cases, the values predicted by the models are better at approximating the runtime than the \textit{time\_limit} value provided by the user. In particular, when the models overestimate the runtime (on average around 80\% of total cases), this results in almost a 98\% improvement (on average). On the contrary, the models underestimate the runtime on average around 20\% of total cases, while the \textit{time\_limit} value does so in just 1.4\% of cases. In about 85\% of these "underestimation" cases, a simple solution could be to add 60 minutes to the predicted runtime, reducing the number of jobs that would be interrupted before finishing to less than 3\% of the total (which is still a lot but more in line with the original value of 1.4\%). Using this "safe" prediction, we have that the models overestimate the runtime 97\% of the time (obviously with higher average error). However, we still have a more than 91\% improvement (on average).

\section{\sched{} Scheduler}\label{sec:sched}

{In this section, we start by briefly describing the logic of the scheduling algorithm built on top of the previous analysis; then we evaluate its performance w.r.t. a widely adopted workload scheduler.}

\subsection{The Scheduling Algorithm}\label{sec:sec:sched_model}

Given the results of the previous section, we propose to enrich the scheduling decisions of an EASY backfilling algorithm with the runtime estimations derived through ML. In practice, the \sched{} algorithm prioritizes the jobs with shorter predicted execution times by operating in the following way:
\begin{enumerate}
    \item It starts by loading historical job data and training a runtime prediction model consisting of a Decision Tree Regressor. This is done only once at the beginning of the algorithm execution.
    \item The runtime of each job is then predicted upon submission, and the time requested by each job is set to this value.
    \item The submitted jobs are sorted based on the requested time and those with smaller predicted runtimes are given higher priority. 
\end{enumerate}
We implemented the aforementioned steps into Batsim \cite{batsim_jsspp16}, an infrastructure simulator that allows the development and testing of resource management policies. The code of our Batsim-based \sched{} implementation is publicly available on GitHub\footnote{URL redacted due to blind policy - it will be made public in case of acceptance}. The repository also includes all the tests reported in this paper and the setups to reproduce the experiments.

\subsection{Experimental Evaluation}\label{sec:sec:sched_results}
In order to evaluate the performance of \sched{}, we start from the dataset of job runs used in Section \ref{sec:time_prediction_result}, which consists of almost 630'000 rows, and we divide it into two parts: 
\begin{inlinelist}
    \item \textit{df\_sched}, which contains the data relative to the last 24 hours stored in the original dataset (a total of 4’407 jobs), and is used to instruct Batsim about the amount of resources and execution time that each job will need at simulation time; 
    \item \textit{df\_train}, which contains the rest of the data, and is used to train the regressor.
\end{inlinelist}

For our experimental evaluation, we wanted to maximally highlight the contribution obtainable by incorporating the duration prediction into the scheduling policy. Hence, we opted for the classical EASY backfilling algorithm \cite{wong2007evaluating}, which we will dub \easybf{} from now on, as the baseline. As previously underlined, \easybf{} is a relatively simple, but still widely used.

The simulation was first carried out on a Batsim platform with a total of 15,680 computing resources, i.e., equivalent to what is available on the MARCONI100 infrastructure (980 physical nodes with 16 cores each). In the following, we refer to this configuration with \textit{Setup A}. Then, aiming to test the schedulers' performance in stressing conditions, we repeat the experiments with a more constrained platform, consisting of just 512 computing resources (\textit{Setup B}). 

We compare \sched{} and \easybf{} based on the following values emerging from the simulations:

\begin{inlinelist}
    \item \textit{makespan}, is the completion time of the last job;
    \item \textit{scheduling time} is the time (in seconds) spent in the scheduler;
    \item \textit{mean waiting time} is the average waiting time observed on jobs, intending it as the time between job submission and its actual start time. It corresponds to the amount of time a job spends waiting in the queue before it starts executing. 
    \item \textit{mean turnaround time} is the average turnaround time observed on jobs, intending it as the difference between the time instant in which the job ends and the submission instant. Hence, the turnaround time includes both the time spent waiting in the queue and the execution. It reflects the efficiency of the system in handling jobs.
    \item \textit{mean slowdown} is the average slowdown observed on jobs. The slowdown of a job is useful for understanding how scheduling affects the performance of individual jobs because it measures how much longer a job takes to complete compared to its actual execution time. It is computed as: ${slowdown} = \frac{{turnaround\_time}}{{execution\_time}}$;
    \item \textit{maximum waiting time} is the maximum waiting time observed on a job;
    \item \textit{maximum turnaround time} is the maximum turnaround time observed on a job;
    \item \textit{maximum slowdown} is the maximum slowdown observed on a job.
\end{inlinelist}

Table \ref{tab:24big} refers to \textit{Setup A} and reports the values of these metrics for both \sched{} and \easybf{} schedulers. 
\begin{table*}[]
    \centering
    \begin{tabular}{|l|c|c|c|}
         \hline &  \sched{} & \easybf{} & {Improvement}\\\hline
         \textit{makespan} & 86272.0068 & 86272.0024 & +0.00\% \\
         \textit{scheduling time} & 37.8449 & 240.7396 & -84.28\%\\
         \textit{mean waiting time} & 846.5391 & 953.3813 & -11.21\%\\
         \textit{mean turnaround time} & 2351.1828 & 2458.0250 & -4.35\%\\
         \textit{mean slowdown} & 2.3089 &45.8519&-94.96\%\\
         \textit{max waiting time} & 17003.0928 &12608.0384&+34.88\%\\
         \textit{max turnaround time} & 64657.0068 & 00657.0024&+0.00\%\\
         \textit{max slowdown} & 261.0818& 12156.04.06& -97.85\%\\
         \hline
    \end{tabular}
    \caption{Comparison of \sched{} and \easybf{} performance when scheduling jobs on a large HPC system (\textit{Setup A}). Negative values in the ``Improvement'' column highlight desirable situations where \sched{} brings a decrease in the corresponding metric.}
    \label{tab:24big}
\end{table*}
Observing these results, we can highlight that \sched{} brings some clear improvements over \easybf{} scheduler. The most relevant is the fact that with \sched{} the mean waiting time of a job is more than 11\% lower. Also, the mean slowdown is significantly improved (-94.96\%).

On the other hand, using the \sched{} scheduler, the maximum waiting time is higher than that obtained using the \easybf{} scheduler by a significant margin (almost 35\%). Indeed, as \sched{} is better at estimating the jobs' duration beforehand, it is also able to identify how a few jobs are extremely more time-consuming than others and, accordingly, it changes their position further down the queue. 

It is worth noticing that when using \sched{}, the waiting time is very low (less than 1 minute) for 4.28\% more jobs.
Observing these results, we can highlight that \sched{} brings some clear improvements over \easybf{} scheduler.
Besides testing \sched{} on a Batsim, large infrastructure, we want to analyse how it performs on an extremely constrained system such as the one in \textit{Setup B}, where a limited amount of computing resources is made available to jobs. 
Table \ref{tab:24small} reports the metrics values for this case. 
\begin{table*}[]
    \centering
    \scriptsize
    \begin{tabular}{|l|c|c|c|}
         \hline &  \sched{} & \easybf{} & Improvement\\\hline
         \textit{makespan} & 1029198.2116 & 1090869.2938 & -5.99\% \\
         \textit{scheduling time} & 210.3460 & 194.5042 & +7.53\%\\
         \textit{mean waiting time} & 127474.5570 & 163846.3107 & -28.54\%\\
         \textit{mean turnaround time} & 128979.2008 & 165350.9545 & -28.21\%\\
         \textit{mean slowdown} & 22785.2399 & 20097.1711 & +11.80\%\\
         \textit{max waiting time} & 994211.2116 & 1026349.2598 &-3.21\%\\
         \textit{max turnaround time} & 1024094.2116 & 1027933.2894 & -0.37\%\\
         \textit{max slowdown} & 450511.6491 & 1026350.2601& -128.09\%\\
         \hline
    \end{tabular}
    \caption{Comparison of \sched{} and \easybf{} performance when scheduling jobs on a constrained HPC system (\textit{Setup B}).}
    \label{tab:24small}
\end{table*}

\begin{figure}[htbp]
    \centering
    \begin{subfigure}[b]{0.45\textwidth}
        \centering
        \includegraphics[width=\textwidth]{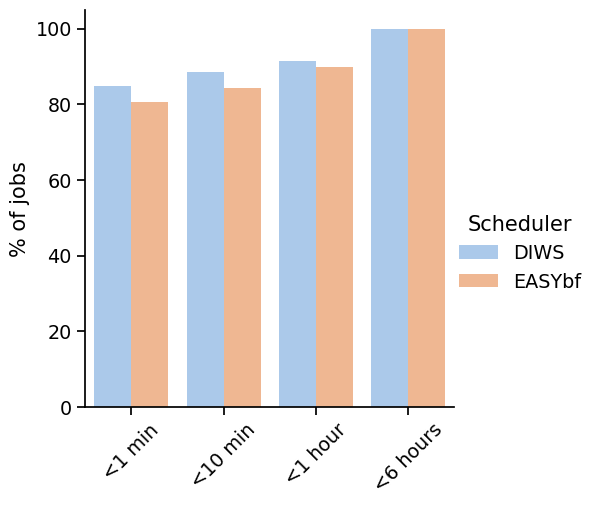} 
        \caption{\textit{Setup A}}
        \label{fig:24big}
    \end{subfigure}
    \hfill
    \begin{subfigure}[b]{0.45\textwidth}
        \centering
        \includegraphics[width=\textwidth]{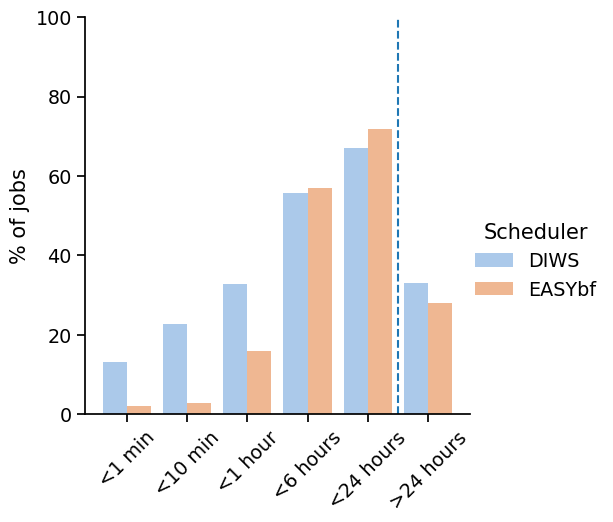} 
        \caption{\textit{Setup B}}
        \label{fig:24small}
    \end{subfigure}
    \caption{Scheduling performance comparison}
    \label{fig:sched-perf-cmp}
\end{figure}

The histograms in Fig. \ref{fig:sched-perf-cmp} show a comparison of the percentage of jobs that wait less than arbitrarily chosen time intervals (one minute, ten minutes, one hour and six hours), for the \textit{Setup A} and \textit{Setup B}.

\sched{} shows the best performance in this constrained setup too. In particular, (from Table \ref{tab:24small}) the total time required to go through all jobs in the workload is almost 6\% lower with \sched{} to \easybf{}, the mean waiting time of a job is more than 28\% lower, and (as shown in Fig. \ref{fig:24small}) the waiting time is less than 10 minutes for almost 8 times more jobs (going from 2.06\% of \easybf{} to 22.76\% of \sched{}).

On the other hand, when using the \sched{}, the waiting time is very high (more than 1 day) for almost 5\% more jobs than when using the \easybf{} scheduler (1457 vs 1241). This is a direct consequence of the very constrained testing environment and--as already pointed out for \textit{Setup A}--the better capacity of \sched{} to estimate the durations, highlighting the huge difference existing between jobs.

\section{Conclusion}\label{sec:conclusion}
As ML techniques have shown promising results in several scientific fields, we propose to apply analogous methods to the HPC scheduling too. Our preliminary analysis takes into consideration four different ML techniques and analyses their ability to predict the execution time of HPC jobs before their submission. The tests, conducted on an extensive real-life dataset of job runs, clearly show the enhancement that ML can bring to runtime prediction provided by users.
Furthermore, we employ a well-known HPC workload simulator to evaluate the efficacy of a duration-informed scheduler by comparing it with a widely used alternative. The proposed solution's clear superiority when aiming to reduce the average waiting time. 

\section*{Acknowledgements}
This work has been partially supported by European Project HORIZON-EUROHPC-JU-SEANERGYS (g.a. 101177590). 

\bibliographystyle{splncs04}

\bibliography{bib}

\end{document}